# Apuntes sobre teoría del comportamiento corrupto: nociones cibernéticas e informáticas para una actualización de la ecuación de Klitgaard[1]

Rodrigo Lopez-Pablos[2]

Febrero de 2015

(Versión en formato ecológico)

**Clasificación ACM:** J.4; K.4.1
**Clasificación JEL:** D73; D80

## 1. Liminar

La sociedad informacional contemporánea posee características de dificultad y problemáticas sistémicas sin precedentes consecuencia de los saltos de complejidad que permitieron las tecnologías de la información; a diferencia de antes, siendo que toda sociedad humana resguardo siempre una complejidad inherente a su ser, ahora la multiplicidad informativa amplifican tal complejidad ilimitadamente puesto que el Ciudadano consumidor y productor de información se ve forzado a desarrollar una heurística básica para comprender la sociedad en la que se encuentra inserto. Los sistemas de información, la humanidad y el fenómeno corrupto inserto en los mismos procesos no son ajenos a esta transformación en ningún aspecto.

Sin embargo así como lo es el tiempo para el individuo biológico, las capacidades biológicas del ser humano para procesar información son también limitadas, especialmente para el Ciudadano común que además del esfuerzo volcado para lograr su subsistencia también sueña con una participación cívica, una ejemplaridad propia y una sociedad mejor para sí y para su trascendencia en la forma de su próxima generación de descendientes. Esto no representa una tarea fácil puesto que estos deben lidiar con la comprensión de sistemas sociales cada vez más complejos dado que la sociedad digitalizada en donde la dependencia entre las funciones de sus componentes es cada vez mayor así como lo es su *interdefinibilidad* lo que caracteriza a todo sistema complejo (García, [9]).

Paralelamente a tal complejización societaria, la evolución de la herramienta tecnológica como medio de transformación social también sufrió saltos de complejidad como parte de ese mismo proceso, igualmente a la aparición de la cámara de video como instrumento de observación del fenómeno corruptivo puntual de agentes públicos y privados, se abren nuevos paradigma a una escala tan significativa como igual de imperceptible en lo que hace al advenimiento de las tecnologías de la información, más precisamente, la de técnicas de explotación de la información y minería de datos sobre todo tipo de bases públicas y privadas, esto como una parte innegable de una revolución en los datos públicos en todos los ámbitos sociales.

Estos nuevos escenarios de complejización sistémica entorno al fenómeno corrupto exigen de nuevas estructuras epistémicas para su comprensión y análisis para las capacidades humanas cada vez más limitadas para procesar información y controlar el comportamiento corrupto de sus dirigentes en un accionar Ciudadano, estos a su vez con una existencia digital cada vez más pronunciada. Este trabajo aborda esta cuestión desde la necesidad de incorporar conceptos de teoría de la información, cibernética comunicacional al tratamiento convencional existente en teoría de la corrupción a las problemáticas informacionales actuales para el desarrollo y mantenimiento de una planificación generacional fenoménica desde el Estado y el Ciudadano como se presenta subsiguientemente con mas detalle.

---

1. Estos apuntes surgen de investigaciones sobre informática aplicada al fortalecimiento del contralor civil ciudadano en la lucha anti-corrupción desarrollado a lo largo de 2014. Todos los hipertextos referenciales disponibles se encuentran validados a la fecha.
2. Investigador en ciencias económicas e informáticas UNLaM y UTN. Usuales cláusulas de responsabilidad se aplican.

Después de describir y proponer una expansión útil para la ecuación de corrupción de Klitgaard en la siguiente sección, se estudia a posteriori los vínculos entre la densidad del capital societario y la existencia de equilibrios éticos ejemplificativos y corruptivos en la sociedad. Se discute la utilidad de los índices de percepción como medio de lucha contra el fenómeno, y la información como objeto corruptivo cuando esta es distorsionada tanto corporativamente como por fricciones dentro del propio Estado. Luego se considera la información como objeto corruptivo *per se*, cuando el ciclo de retroalimentación informativa cibernética entre el Estado y su ciudadanía se ve corrompida al interrumpirse los servo-mecanismos comunicacionales de autoregulación societal.

## 2. La ecuación de Klitgaard y sus variantes

La notable ecuación sobre la explicación del comportamiento corrupto desarrollada por Klitgaard [11] propone la explicación del hecho corruptivo a partir de una simple ecuación, frente a la profundización de los procesos informacionales actuales, abre la posibilidad a una actualización de acuerdo al presente donde la aceleración de la tecnologías informáticas e comunicacionales potencian una revolución digital en proceso.

Siendo que el desafío no es solo proponer un formato para su actualización sino mantener su simplicidad, para comprender su carácter, la ecuación originaria de Klitgaard puede describirse en su versión originaria de la siguiente forma.

$$C = M + DR - TR \qquad (1)$$

Donde el nivel sistémico de corrupción de una sociedad está representado por «C», el cual, como se desprende dependerá directamente del poder de los funcionarios públicos (M) así como del grado de discrecionalidad de las normas y leyes regulatorias para la rendición de cuentas por parte de los organismos de control de Estado, la sociedad civil y el periodismo crítico (DR) e inversamente proporcional a la transparencia y responsabilidad de los mandantes (TR) A partir de la la **ecuación** 1 su pedagógica correspondencia semántica entendida como:

«Corrupción es igual a monopolio más Discrecionalidad (Regulatoria) menos Responsabilidad (Transparencia)»

Esta es tan simple y divulgable para los expertos del campo del arte que Klitgaard la concibió como la *fórmula de la corrupción*; consecuentemente para Klitgaard, la conducta ilícita siempre florecerá siempre que los actores con poder de algún tipo tengan monopolio, poco control y poca transparencia.

A partir del primer esbozo en la **ecuación** 1, podemos afirmar que Klitgaard, al intentar interpretar el fenómeno corruptivo consideraba al mismo como un fenómeno sino complejo eminentemente multidimensional coincidiendo con otros estudiosos del campo (Moreno Ocampo, Raigorodsky y Geler; [15], [19]).

Otro aspecto ineludible es el de la realidad informática de tal contemporaneidad, puesto que la concentración mediática en la forma de medios tradicionales presupone otra omisión imposible de eludir, siendo sus flujos y canales de información otro elemento de poder capaz de ser objeto de planificación social sin control sobre la sociedad receptora de información cuando aquella carece de una ética mínima hacia las generaciones del presente y del futuro. De esta forma, cuando sus fuentes no poseen una originalidad democrática auténtica, y un nivel de monopolización considerable, sus emisiones pueden ser suficientes para generar rupturas en el tejido y la cohesión social, inestabilidad social que coarta el bienestar de las poblaciones afectadas.

No es la primera vez que se intenta extraer conocimiento de una variación de la fórmula de la corrupción de Klitgaard, Moreno Ocampo [15] ya lo había intentado buscando identificar la vinculación existente entre la corrupción y la organización política del Estado, a partir de la misma ecuación inicial dedujo la siguiente expresión:

$$PE = M + DR - TR \qquad (2)$$

Este, a partir de los conceptos de la **ecuación** 1, Moreno Ocampo reemplaza el concepto de corrupción por el de poder hegemónico (PE) sin mayor profundidad alguna, emulando sin más la correspondencia semántica de la **ecuación** 1 esta se traduciría ahora de la siguiente forma:

«poder hegemónico es igual a monopolio mas discrecionalidad menos transparencia»

En base a esta proposición, Moreno Ocampo [15] termina concluyendo que Democracia (Dem) es distinto a poder hegemónico y por ende distinta a la definición de corrupción de Klitgaard tal que:

$$PE = Dem \qquad (3)$$

Aunque presenta un precedente y pudiésemos afirmar que más democracia constituye menos corrupción, esta podría pecar por ser una deducción demasiado simple y conceptualmente demasiado amplia de manera que pudiese ofrecer una solución adicional a un problema siempre complejo como lo es el de la corrupción en las sociedades modernas. Por ello se propone robustecer la ecuación incorporando ideas se representación social como lo son el capital social y nociones relativas a los entornos informáticos, cibernéticos y digitales actuales como se ofrece a continuación.

## 3. Incorporación del capital societario y componentes informacionales

Ya interiorizados de los fundamentos básicos de la ecuación, se procede a la integración de la misma asumiendo que el ilícito donde se produce el hecho corruptivo son sociedades que poseen un capital societario determinado, así como se encuentran insertas en entornos digitales propios de contextos actuales. En su versión expandida la ecuación de Klitgaard queda de la siguiente manera:

$$C = M + DR - TR \pm DCS + BD - NR - RI + IRC \qquad (4)$$

En la **ecuación** 4 expandida se aprecia la incorporación a la formula de la densidad del capital social o societario (DCS), la brecha digital efectiva existente en la sociedad (BD), la neutralidad de la red (NR), el grado de discrecionalidad de la regulación informática (DRI), y el efecto de la complejidad sistémica societaria a través del ruido e interferencia comunicacional (IRC).

Si por ejemplo la regulación informática es inefectiva o se encuentra descoordinada de la estrategia generacional del Estado se abre la posibilidad a la aparición contingente de interferencia aleatoria dentro de la sociedad toda. Lo que termina en desinformación, angustia innecesaria y derroche energético cognoscitivo y temporal de la población[3].

El IRC quizás sea el componente –en términos de sistemas complejos– más *interdefinible* y sistémico puesto que es a través de este donde las perturbaciones societales intervienen absorbiendo tiempo de observación irreversible del Ciudadano así como el componente no religable de mayor impacto sobre un capital social determinado como se verá más adelante. De manera inversa por otra parte este es el reflejo del grado de retroalimentación y (des)equilibrio cibernético de la propia sociedad y el rol que cumplen los medios de masas servomecanismos de auto-control societario en la sociedad (Chu, [6]).

Desde esta perspectiva cibernética informacional una sociedad que cuente con un capital societario más consolidado lo será también parte de una sociedad más estable informacionalmente, de manera de poseer medios informativos que cumplan un rol de servomecanismo de autoregulación en la relación de los Ciudadanos con el Estado; consecuentemente, es útil presuponer que existe un (des)equilibrio ético que sostiene la densidad del capital societario, como pilar de sustento del bienestar, este debe ser apuntalado informativamente.

Además del anterior una versión expandida de la ecuación además de los términos propuestos por Klitgaard debería reflejar las relaciones entre sus componentes como se despliega en la siguiente forma.

$$C = M + DR \varnothing TR(IC, AEIP) \pm DCS(EE^*/EC^*) + BD(UC, CC) - NR - RI + IRC(PI) \qquad (5)$$

En donde además de la configuración de la **ecuación** 4 (expandida), se expone el funcionamiento de algunos de sus componentes y las relaciones contenidas dentro de cada una: mientras el nivel de transparencia y responsabilidad ética de los oficiales públicos dependerá en la existencia y efectividad de instituciones de contralor anti-corrupción (IC) así como del acceso efectivo a la información pública (AEIP), la profundidad de la brecha digital lo hará del nivel de universalidad en la conectividad poblacional (UC) y la (auto)participación ciudadana en el contralor civil (CC). La densidad del capital societario obedece a la existencia de un equilibrio ético societario en el cual descansan los servomecanismos cibernéticos informacionales de auto-regulación societal; según la fase dialélica positiva o negativa que la sociedad este atravesando esta podrá ser correspondiente a la de un equilibrio ético corruptivo (EC) o la de un equilibrio ético ejemplificativo (EE).

La interferencia y ruido comunicacional en una sociedad se ve relacionado a la presión informativa (PI) a la que se encuentra sometida, –*e. g.* cuando medios con capacidad de difusión masiva intervienen destructivamente los lazos informacionales entre el Estado y el Ciudadano coartando sus horizontes planificativos o cuando un dirigente político corrompe sus principios partidarios difundiendo a sus militantes otro mensaje distinto al de su formación originaria, etc.– la información como recurso útil para la planificación del individuo fenoménico se ve corrompida impactando a nivel agregado en Estado benefactor plasmado en la aseguranza de un capital social establecido, capaz de acortar las desigualdades sociales mediante planificación generacional contra la desigualdad y la pobreza estructural.

Desde la centralidad que posee el capital societario en los niveles de corrupción y desarrollo de toda sociedad, un escenario de equilibrio ético ejemplificativo estaría asociado naturalmente a un dialelo positivo a sociedades menos corruptas, donde las estrategias corruptivas no sean las dominantes, no por imposición gubernamental sino por un servomecanismo automático societal asociada al bienestar y un Estado planificativo benevolente efectivo.

---

3. Un ejemplo de ello puede apreciarse en el caso de excesiva autonomía operacional y descontrol del aparato de inteligencia Argentino, desencajado este de toda política de Estado, no sólo no rendía cuentas a la ciudadanía sino que además era capaz de condicionar tácitamente a organismos democráticos de la república (ADC, [1]).

Por otro lado, la noción de un equilibrio corruptivo se relaciona más con la de un dialelo negativo, cuando la corrupción erosiona la confianza en las instituciones del Estado, lo que a su vez debilita la capacidad del Estado de luchar contra la corrupción (Bour, [2]), esta puede ser atribuida a un Estado pasivo en retirada que ha fracasado planificativamente en proteger el bienestar social al fundamentar sus decisiones en teoría e información falaz dentro de una fase descendente de destrucción del capital societal o vinculado a una sociedad anómica. El mismo autor propone reformas sistemáticas concentradas en todo el sistema desde una perspectiva economicista afirmando que una mayor liberalización económica reducirá la corrupción, cuando ya han sido comprobados las nefastas consecuencias que posee el neoliberalismo sobre el capital social, su descomposición cultural e histórica, en sociedades de economías en desarrollo periféricas.

En un Estado planificador benevolente en cambio, satisfactor de las necesidades poblacionales, es más probable que este se alimente un circulo virtuoso en su función anti-corruptiva puesto que con su accionar soporta un mayor capital social para una mayor proporción de ciudadanía lo que termina retroalimentándose positivamente en el fortalecimiento del capital social en una mayor auditoría cívica –mediante mayor participación ciudadana–, mayor discusión pública sobre el fenómeno, así como se formaran lideres políticos éticamente responsables.

De esta forma el componente central quizás de mayor importancia en la ecuación para el bienestar generacional de la población viene a ser el de la densidad del capital societario (DCS). La explicación del mismo a partir de la **ecuación** 5 viene a expresarse de la siguiente manera:

$$DCS(EC^*/EE^*) = \pm C - M - D + R(IC, AIPE) - BD(UC, CC) + NR + RI - IRC(PI) \qquad (6)$$

Del cual dependerá centralmente si dentro del Estado asegurador del bienestar societario se encuentra dentro de un Equilibrio corruptivo o ejemplificativo, para que a partir del equilibrio dado el comportamiento corrupto contribuya o no a la densidad del mismo. Paralelamente menor discreción, mayor regulación anti-corrupción, una menor brecha digital, mayor neutralidad en la red contribuirán a una mayor densidad del capital societario independientemente del equilibrio ético que prime en la sociedad.

De esta manera un Estado con responsabilidad informativa, a sabiendas del impacto sobre la densidad del capital societario de su pueblo, promoverá la inclusión digital de la ciudadanía que se vea imposibilitada de acceder a las tecnologías digitales a través de su universalización al acceso con carácter de servicio público, así como regulará la concentración de los flujos informativos, atomizando la oferta cultural fundamental para la formación cívico-política digital del presente. La corrupción por otra parte tendrá un efecto ambiguo sobre el capital societario dependiendo en que tipo de equilibrio ético se encuentre la sociedad, ya que aquella será efectiva para sostener la densidad cuando el equilibrio sea corruptivo e inefectiva cuando el último sea ejemplificativo.

# 4. Los equilibrios éticos fenoménicos corruptivo y ejemplificativo en un modelo de mecanismos informativos

Además del necesario intento de incorporar nociones informativas y digitales al estudio del comportamiento corrupto, estos también deben aproximarse al estudio del fenómeno de manera dinámica, algo que Klitgaard deja de lado al menos en lo propuesto en su ecuación explicativa fundamental. En su forma expandida se intenta incorporar la noción de un equilibrio ético dentro del componente central de densidad del capital societario el cual puede ser de caracter corruptivo o ejemplificativo.

La importancia de la densidad del capital societario en la población es clave pues no solo determina su bienestar sino la mayoría de los fenómenos societales que provienen de este, entre ellas el del fenómeno de la corrupción; de ahí la importancia de su comprensión, de sus estados de equilibrio o desequilibrio sociocultural, las éticas mínimas presentes que aseguran de cohesión social, el devenir del fenómeno corruptivo o ejemplificativo dentro del grupo, puesto que son esas ejemplificaciones locales en la forma de Ciudadanos solidarios y estos como los principales garantes del cumplimiento de las leyes, de la acción de la justicia y de los gobiernos, siendo –a partir de estos– el gran desafío que tiene toda sociedad el de afrontar la corrupción (Marin, [17]).

La existencia de un *equilibrio de corrupción* fue explorado a través de teoría de los juegos como se propuso en un modelado propuesto de Bour [2], como comúnmente se propone con buenos resultados didácticos y útil para el conocedor de la teoría, este empero, deja de lado la complejidad inherente del fenómeno corrupto, el cual va más allá del abordaje netamente economicista estándar, al no incluir por ejemplo aspectos como los del capital social. No obstante, este puede ser enriquecido si se plantean los supuestos de un jugador que planifica generacionalmente o que en otras palabras, tiene plena conciencia de la idea de la muerte y su finitud biológica en este mundo, *i. e.* un planificador fenoménico auténtico.

Dentro de un equilibrio ético-corruptivo los mecanismos informacionales repercutirán al menos de dos maneras, (i) retrasando el tiempo para la creación de consensos en la *espiral del silencio*[4] lo que a su vez ralentiza la reacción societaria ante amenazas externas sobre su cultura originaria; (ii) al contar con información difusa los prospectos fenoménicos de decisión del actor, su entorno y por ende sobre el bienestar fenoménico de toda su sociedad, lo que a su vez implica que la sociedad afectada descansen sus prospectos de proyectos en un equilibrio de corrupción en lugar de un equilibrio de ejemplaridad como se desprenderá de un ejemplo vertido más adelante.

Adoptando una abstracción similar a la desarrollada hace unos años (Lopez-Pablos, [13]), se asume la existencia de un mecanismo informacional sobre dos individuos fenoménicos ambos pertenecientes a un mismo territorio y una misma génesis cultural, ambos, a pesar de acceder a las mismas fuentes emisoras de información poseen naturalmente distintos prospectos de proyectos fenoménicos, aunque diferentes ambos esperan transmitir su cultura a su próxima generación de descendencia para trascender a sí mismos; por lo tanto, ambos deben ser ejemplares en su accionar dentro de los contextos en los que se sitúan. Sin embargo, la planificación de sus prospectos les exige que haya una transacción entre ellos en un segmento del tiempo, y la situación informativa y societal de cada uno.

Esta vez a diferencia de otras abstracciones para el estudio del comportamiento corrupto donde se adopta la representación de un agente privado y un funcionario público, se supone ahora la oportunidad de corromper o ejemplificar la norma entre dos funcionarios públicos, ambos oficiales del Estado con distintos desempeños y áreas de influencia, lo que en caso de haber ruptura de la norma se estaría recayendo en perjuicio social doblemente agravado al tratarse de actores necesarios para la articulación con la gobernanza, por ende resultando en el peor tipo de deshonestidad hacia el propio tejido social la corrupción intra-Estatal entre responsables oficiales. Naturalmente, en cuanto mayor sea el grado de responsabilidad de los oficiales corruptos involucrados mayor será el daño informacional derramado sobre el capital social, así como mayor

---

4. Concepto proveniente de teoría de la comunicación, entendida la opinión pública como un medio-sistema para mantener la cohesión societaria en la que los individuos adaptan su comportamiento a las actitudes predominantes sobre lo que es aceptable o no, los medios masivos pueden optar por amplificar la conducta ejemplar o la corrupta, amplificando en la acción comunicativa el lado moral o inmoral de cada fenómeno social (re)transmitido.

será su fortalecimiento cuando entre ambos se adopte una postura ejemplar.

**Ejemplo 1** *Adoptando una ejemplificación abstracta* X = [*Corromper; Ejemplificar*] *con* I = 2 *con dos simples prospectos de proyectos generacionales con distinta densidad de capital social cada uno sujeto a distintas fuentes informativas, tal que:* $\Theta_1 = [\theta'_1; \theta''_1]$ *sujeta a una fuente informativa mono u oligopolizora del tiempo con escasa ejemplaridad de los líderes políticas,* $\Theta_2 = [\theta'_2; \theta''_2]$ *sujetas a una fuentes informativas diversas con gran ejemplaridad política, las posibles preferencia de los agentes planificadores serán ergo* $S_1 = [\succcurlyeq'_1 (\theta'_1); \succcurlyeq''_1 (\theta''_1)]$; $S_2 = [\succcurlyeq'_2 (\theta'_2); \succcurlyeq''_2 (\theta''_2)]$ *t. q.:*

| **Prospecto I ($\theta'_i$)** | | **Prospecto II ($\theta''_i$)** | |
|---|---|---|---|
| Natalio | Maria | Natalio | Maria |
| Corromper | Corromper | Ejemplificar | Ejemplificar |
| Ejemplificar | Ejemplificar | Corromper | Corromper |
| Corrompe óptimo (EC*) | | Ejemplifica óptimo (EE*) | |

Lo que en términos de la correspondencia de elección social ex-post tendríamos:

$$f(\theta'_i) = \textbf{Corromper} \qquad f(\theta''_i) = \textbf{Ejemplificar}$$

En el mismo se supone la propuesta de dos mecanismos informacionales, uno a través de múltiples emisores y otro a través de uno o muy pocos emisores, en ambos casos luchando entre si por absorber el tiempo de observación del receptor, donde ambos prospectos receptores asignan su tiempo a consumir distintos canales informacionales, ambas estrategias útiles para la realización de sus prospectos fenoménicos en función de la autenticidad o no de la usina informacional predominante.

Como propone el modelo teórico, en el primer prospecto contempla el contrato propuesto por Natalio considerando un contexto informativo de consumo monodimensional o multidimensional donde, en base a este y la densidad del capital social de la comunidad al que pertenece, sopesa la decisión de corromper el sistema o ejemplificar sobre el mismo de acuerdo a la información provista por el mecanismo. Dada la propuesta del primer jugador, Maria deberá optar, contando solo con la información provista por la monousina informacional, en aceptar o no, y consecuentemente en corromper o ejemplificar con su accionar al sistema social al que ambos pertenecen; en cualquiera de los casos, sus decisiones generarán antecedentes y esto mismo será parte del acervo informativos para la siguiente generación de descendientes culturales y biológicos; *i. e.* conjuntamente son conscientes que ellos mismos generan un derrame informacional positivo o negativo en su mismo capital societario originario.

Dada su irreversibilidad, ambos son conscientes del peso decisional respecto el tiempo sus decisiones consumen todas sus energías vitales, dentro del **prospecto I** el jugador se encuentra mas propenso a ser manipulado por el generador del mecanismo monoinformacional, más distanciado del capital societario que él mismo compone y observando parcialmente la ejemplaridad de sus líderes; es más probable ergo que elija corromper al menos para forzar una respuesta dentro de su situación en el propio sistema y obtener información a partir de este. Por otro lado, Maria, deberá evaluar la decisión de corromper con las mismas fuente y su propia percepción sobre su capital societario, de otra forma optará por ejemplificar.

Naturalmente al contar con la creencia completa sobre solo una fuente emisora la probabilidad de desinformación es mayor, lo que tendería a desintegrar el capital societario hacia un punto crítico revolutivo; ergo, si ambos hubieran optado por corromper como su óptimo, se estaría verificando la hipótesis de que la sociedad a la que ambos pertenecen sufre de corrupción estructural generalizada; de otra forma, con decisiones disímiles la sociedad a la que pertenecen Natalio y Maria se encontraría en algún punto medio entre corrupción estructural y marginal.

Otro sería el caso si se presupusiera un mecanismo informativo diverso, lo que a su vez provee una apreciación más completa del capital societario compartido entre ambos, con consensos y discusiones públicas más desarrolladas sobre temas públicos, un lenguaje intrasocietal más claro y con mayor ejemplaridad ciudadana y política –como bien se propone en el **prospecto II**–. Aquí siendo que ambos encontrasen como óptimo ejemplificar, su capital societario se fortalecería en un dialelo positivo, estancándose este en caso que la elección de uno fuese disímil y colocando también a la sociedad en otro punto medio entre corrupción estructural y marginal.

El tiempo y la información son los insumos básicos de estos prospectos, ante la restricción temporal universal de la finitud de la vida humana como constante ontológica única, el tiempo sobreviene en el recurso más escaso y valioso del individuo así como al consumo de desinformación al derroche y despilfarro irreversible del mismo, lo que termina repercutiendo en el fracaso de prospectos auténticos.

En caso de paridad en una comunidad con diversidad de usinas mediáticas en ambos prospectos, será el grado de ejemplaridad de los líderes de la comunidad representado lo que determine la probidad de la misma, y por ende, que el equilibrio de la prole sea corromper o ejemplificar. Es a la vez esta ejemplaridad de los mismos lo que sitúa a una sociedad como estructural o marginalmente corrupta, puesto que cuanto mayor fuera la responsabilidad del funcionario corrupto o probo, mayor será el impacto de su accionar sobre su propia comunidad y viceversa.

De esta manera la existencia de un equilibrio ético ejemplificativo o corruptivo en la sociedad, más allá de ayudar a comprender un poco más la compleja trama del fenómeno corruptivo, nos ayuda a rescatarnos no solamente de mejores explicaciones que pueda tener la explicación del fenómeno en su estudio economicista, sociológico o psicopolítico, sino de la importancia central que posee el capital social en el bienestar general de la comunidad y el fenómeno corrupto dentro de ella.

Un caso de estudio interesante de anómia generalizada donde la corrupción se encontraba estructuralmente aceptada puede revisarse en Marin [16], el cual guarda pormenorizado detalle de uno de los tiempos más nefastos de destrucción profunda de capital societario en la sociedad Argentina de fines del siglo XX, en un claro reflejo psicopolítico societario de supervivencia la propia cultura se vuelve transgresora de sus propias normas para mantener la cohesión de su propia estructura. Aquí puede apreciarse como la insensibilidad del propio sistema, sus líderes y todo el sistema republicano representativo se fagocitaba sus propias ejemplaridades vivas más incólumes como se expide Marin con la decisión ejemplificatoria de Rene Favaloro. Es imposible determinar aún cuanto tiempo ahorrado hicieran ganar los baluartes morales a sus sociedades con sus decisiones probas, así como gracias a estos, cuenten ahora con varias manos más en las qué sostener sus proyectos civilizatorios.

## 5. El problema de la medición del fenómeno corrupto

El problema de mensurar de alguna manera la ciudadanía moral de una población no solo se vuelve un problema metodológico al intentar de dimensionar sino el de una necesidad para estudiar positivamente las carencias morales de un grupo humano. Lamentablemente hasta ahora los ejemplos más difundidos mediáticamente descansan en los llamados *índices de percepción de la corrupción*, a pesar del esfuerzo probo a través de asociaciones que luchan contra la corrupción, muchas de ellas se centran en la elaboración de estos índices cuando es sabido que reflejan las percepciones y no una medida cuantitativa y objetiva de la corrupción actual (Tanzi, [22]). El problema no radica solo en la unidimensionalidad que estos plantean sino en que estos mismos contribuyen involuntariamente a mantener el *status quo* en sociedades que sí son afectadas por corrupción estructural generalizada, así como preparar el camino para las que aún no lo son.

Tal fenómeno se retroalimenta esporádicamente y de manera automática con la simple difusión de un solo índice entendido como definitorio lo que hace a la *naturalización* del concepto por parte de grandes masas de Ciudadanos que destinan grandes esfuerzos a satisfacer sus necesidades básicas y por ende dejan de destinar energías a la solidaridad que supone la lucha cívica y política. Esto no solo termina en la delegación del ejemplo cívico hacia una misma clase política cada vez más inamovible sino también en la delegación de tales facultades a las mismas usinas mediáticas tradicionales que pretenden representar la opinión ciudadana muchas veces sin filosofía moral alguna.

En otras palabras la naturalización de la corrupción por parte de la amplificación mediática tiene el efecto profundamente nocivo para la construcción de proyectos colectivos al considerar a la corrupción como algo dado, sin posibilidad de cambio y transformación desde abajo hacia arriba. Siguiendo a Rodriguez Kauth y Di Dio [7] los índices de percepción de la corrupción cuentan con las siguientes falencias:

(i) Naturaleza subjetiva del índice.
(ii) Presentan problemas con las unidades muestrales que conforman a los encuestados.
(iii) No se precisa la metodología aplicada, las unidades muestrales del índice de percepción constituyen solamente empresarios y analistas de riesgos sin contemplar la opinión de otros actores.

Siguiendo al mismo autor, los índices de percepción fracasan además en su propuesta de ser un instrumento noble de lucha contra la corrupción al proponer este una naturalización del fenómeno en las sociedades percibidas como corruptas, pues presentan la percepción de un problema pero ninguna solución integral a la misma.

Algo que no puede percibirse a nivel institucional de las asociaciones que los confeccionan pero si cuando aquellos son amplificados por los canales sistémicos a través de los cuales se nutre informativamente la población que intenta ser representada. Cuando esto sucede en sociedades con emisores informacionales no atomizados sin regular ocurren al menos dos fenómenos: (i) Interferencia y ruido en la transmisión informacional entre el Estado y sus componentes individuales –lo que en la **ecuación** 4 ampliada vendría a representarse por la componente IRC– que utilizan esos canales para tomar decisiones económicas y políticas para su planificación fenoménica individual y familiar. (ii) Socava la esperanza ejemplificadora que podría auto-proclamarse desde una ciudadanía ética responsable más completa y extendida reflejada en una mayor cantidad de Ciudadanos.

Estos intentan difundir un dato dando a conocer un impacto subjetivo e intersubjetivo que tiene un fenómeno objetivo como la corrupción, este impacto prospera en el desequilibrio del sistema sociocultural (Rodriguez Kauth y Di Dio, [7]); ergo, las medidas subjetivas de percepción corruptiva acaban contribuyendo a una naturalización fatalista del fenómeno corrupto sin ofrecer mayores aprendizajes para la solución de la problemática real, los índices de percepción terminan convirtiéndose en una aproximación unidimensional incapaz de abordar una planificación a la altura del fenómeno complejo que se intenta resolver.

Al introducirse como información útil dentro del ciclo cibernético-comunicativo societal este se amplifica en cuanto a su naturalización y aceptación del acto de corromper como algo dado e inamovible lo que débese contender al fenómeno desde la conciencia civil, algo que expresa claramente Marin [17] cuando afirma que:

*«[...] todo intento de combatir la corrupción debería contemplar una transformación sociocultural en la que la sociedad modifique una actitud fatalista que impregna la representación del fenómeno naturalizado bajo la creencia rectora de su inevitabilidad».*

Lamentablemente, importantes estudios sobre lucha y métodos anti-corruptivos recientes ignoran los efectos de naturalización fundamentando sus hipótesis apoyándose vehementemente en índices de percepción de la corrupción como verdad revelada -*e. g.* los trabajos de Bour [2],[3] y Phelps et al. [18] por ejemplo constituyen arquetipos en este sentido-, sin considerar que para arribar a un mejor equilibrio ético tales índices pudieran ser sesgados e inútiles, además despreciando los efectos amplificantes de estos sobre la sociedad como un efecto adverso sobre la principal arma con la que cuenta la sociedad para erradicar la corrupción que consisten en la ejemplaridad y la ética ciudadana de la propia población.

Desde las entidades dedicadas a la investigación del fenómeno corrupto, es loable la discusión de la filosofía moral dentro de las asociaciones civiles, los cuales suelen presentar índices en lugar de realizar denuncias concretas, pues deberían ser estas las que expone a las corruptelas y los sistemas mafiosos más poderosos, aquellos que más amenazan con la integridad del capital social de los pueblos, pues además de su proba tarea, su labor ejemplificadora debería ser doblemente moral. Así como es deber de todo Ciudadano o institución que se precie de honesta hacer la denuncia pública, identificando nominalmente a la persona física o jurídica en cuestión y acompañando de las pruebas que así lo acrediten en los estrados judiciales correspondientes, de otra forma se trata de simple ocultamiento de posible comisión de delito (Rodriguez Kauth y Di Dio, [7]).

Por un lado, los indicadores de percepción de la corrupción analizan el fenómeno corrupto a nivel país cuando se sabe que su fenomenología responde no sólo a fallas sistémicas en las naciones que lo sufren, sino también a complejas asociaciones ilícitas internacionales que combinan poder financiero, político e informacional coordinado, que retroalimentan sus decisiones de corromper justamente en aquellas sociedades donde los contextos de corrupción se encuentran más naturalizados. Por esos motivos, más allá del grado de inmoralidad en las sociedades débese primero determinarse discretamente si la comunidad en cuestión se sitúa dentro de un escenario de corrupción estructural o en un contexto de corrupción marginal (Moreno Ocampo, [15]).

Por otro lado se carece de un abordaje evaluativo de la corrupción global intra-estatal e intracorporativa puesto que existen sistemas corruptivos internacionales complejos de índole corporativa los cuales no son observables cuando se estudia el fenómeno a nivel país que además de explotar los vacíos institucionales e informacionales para la obtención de réditos políticos y económicos a nivel local, se potencia cuando los sistemas de inteligencia de los propios Estados no se encuentran regulados ni coordinados con políticas estratégicas del propio gobierno representativo en cuestión.

## 6. Interferencia, ruido, presión y corrupción informativa

Así como ocurre en el acto corruptivo convencional la información es también objeto de corrupción cuando esta es interferida premeditadamente con una estrategia tácita de encriptación dirigida del mensaje emitido. Así como el dinero, el poder político y el tiempo la información también constituye un recurso de valor por la utilidad que este posee para la decisión humana; por estos motivos la información puede ser corrompida para provecho particular aprovechando la energía de la fuente emisora más aún cuando esta es pública.

Más allá de las teorizaciones fundamentales presentadas por Shannon y otros teóricos de la información (Shannon, Ungeheuer; [21], [23]) es necesaria la representación de los flujos informacionales a nivel agregado con capacidad de representación genérica de lenguaje natural en la comunicación de masas también requiere la formulación de la representación de un código popular simplificado –el cual es naturalmente percibido como burdo por las poblaciones demandandantes de necesidades más avanzadas y sofisticadas– en la que las masas de Ciudadanos con tiempo y recursos asignados casi totalmente a su subsistencia puedan tener una planificación mínima para sus horizontes de actuación es esencial.

Una fuente de emisión pública cualquiera –ya sea privada o del Estado– emite información con un tipo de encodificación a través de un lenguaje una cultura determinada de manera que esta pueda ser decodificada por una población objetivo a la cual se dirige la información, de la misma forma en que se produce un encriptamiento de datos, cada fuente elige un nivel de complejidad para la encodificación del mensaje emitido; naturalmente cuanto más compleja y menos general sea la codificación más sectárea será la población objetivo y viceversa.

En este sentido la función comunicadora del Estado se entiende como un cripto-sistema cibernético comunicativo social donde este trasmite a sus Ciudadanos los mensajes necesarios para la planificación fenoménica y la satisfacción de necesidades, la codificación de tal información se da a través de la cultura popular y la historia de cada sociedad, la cual utiliza esta misma para decodificarla y tomar sus decisiones.

Siguiendo teoría informática de crypto-sistemas (Rivest et al., [20]) esto puede expresarse formalmente, siendo $M_0$ un mensaje público del Estado este es encodificado a través de una metodología, –un lenguaje, una historia y una cultura– E, así como es decodificada por con esa misma metodología que poseen los receptores, tal que:

$$DP(EP(M_0)) = M_0 = EP(DP(M_0)) \qquad (7)$$

Donde un mensaje informativo público fidedigno encodificado popular y públicamente $M_0$ es trasmitido por el Estado y difusores masivos de comunicación el cual es codificado por la masa receptora consumidora de información necesaria para su subsistencia y trascendencia. Sin presencia de una fuente de ruido, el mensaje emitido encodificado y decodificado públicamente de la **ecuación** 7 resulta ser el mismo al recibido como se plasma en el siguiente.

$$EP(DP(M_0)) = M_0 \qquad (8)$$

Sin embargo la igualdad de la **ecuación** 8 no puede sostenerse cuando existen fuentes de ruido con poder disruptor, *i. e.* fuentes de emisión no legitimadas en cuanto ser poseedoras de una ética comunicacional acorde a la paz necesaria para consolidar la densidad societaria, desde la falta de responsabilidad moral corporativa se busca corromper la emisión informacional legítima por un lado generando ruido para perturbar la contemplación informativa de las masas; y por otro, –aún más importante– absorber tiempo de observación y contemplación de los receptores posicionándose en el medio del ruido generado, y a través de allí insertar su propia encodificación cultural e histórica útil para los principios y fines de la alta planificación corporativa y no la planificación generacional del Estado como planificador supremo representativa de la planificación fenoménica de toda la sociedad agregada.

$$DP(EP(M_0)) \ = M_0 \quad \text{t. q.} \quad \exists \ IS(M_0) = EP(DP(M_1)) \qquad (9)$$

La desigualdad de la **ecuación** 9 representa el quiebre provocado por la irrupción sistémica informacional y de la auto-regulación cibernética societal como consecuencia de la incomunicación entre el Estado y los Ciudadanos que lo constituyen. De allí se desprenden dos mensajes públicos masivamente emitidos uno popular en el origen $M_0$ necesario para el bienestar mayoritario y un mensaje interferido sistémicamente $M_1$ a partir de una fuente de ruido de un interferidor sistémico (IS) que incrementa la presión informacional en la sociedad y coarta la capacidad de autoregulación cibernética de la sociedad.

En caso de que el mensaje $M_0$ llega al receptor agregado societario decodificado correctamente la sociedad se retroalimenta positivamente a través del accionar cívico y ético de sus propios Ciudadanos cuando tanto el Estado como los medios comunicadores de masas contribuyen a esa eticidad comunicativa, y la participación ciudadana, en este accionar entorno a un equilibrio ético ejemplificativo los sistemas comunicacionales informativos actúan como un servomecanismo cibernético que se auto-regula a si mismo (Chu, [6]).

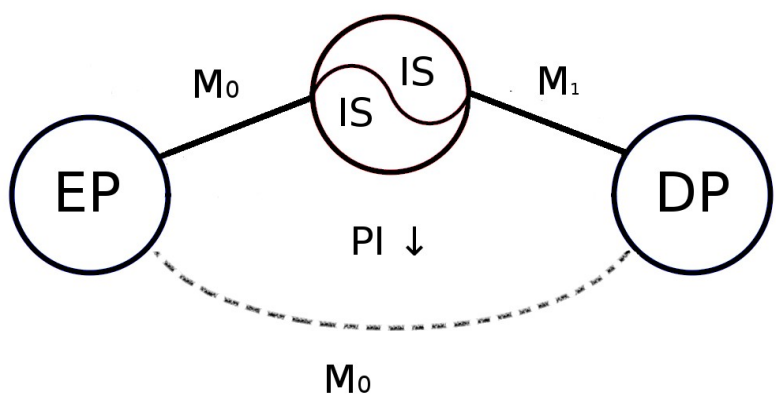

Fig. 1: Interferencia sistemática, presión y corrupción informativa

Como se aprecia del anterior, la fuente de ruido IS crea interferencia sobre los canales del mensaje para luego encodificar la señal a su favor de manera de crear fricción entre los receptores, desgranando la densidad social de los receptores desde el ruido generado.

Esta intromisión muchas veces perjudica enormemente el funcionamiento de la organicidad del Estado dado que algunas de las funciones del mismo consisten en trabajos de investigación los cuales necesitan de un aislamiento societario para poder concentrar energía humana para una función determinada como lo son las causas judiciales, la labor de los organismos anticorrupción de contralor del propio Estado, etc. estos demandan de un ámbito informacional estable y pacífico para acometer su tarea indispensable para la sociedad toda.

Las *funciones* que necesitan de un proceso aislamiento social como los procesos judiciales, la investigación académica, etc no pueden ser popularizadas puesto que necesitan de un tiempo de acumulación cognoscitiva, concentración de energía humana para la generación de conocimiento autónomos, solo en el momento de publicación es cuando se publica y se populariza su labor esta tiene la oportunidad de volverse popular, las interferencias comunicacionales no obstante sobre estos procesos de investigación coartan los objetivos de las mismas así como sus funciones útiles para el Estado y sus Ciudadanos.

Ante estas disrupciones sobre la mecánica informativa del Estado este tiene al menos dos opciones para imponerse cuando estas fuerzas no legítimas intentan doblegar su poder público: (i) Incrementar sus energías comunicativas del Estado en pos de complejizar la codificación pública desde sus representantes a la sociedad; esto es, enfrentándose al peligro que la codificación sea tan compleja que el Ciudadano común no tenga el tiempo y la energía para decodificarlo. (ii) Llegar a la mayor regulación pública posible de la información asegurando la libertad de expresión, la cual podría hacerlo institucionalizando un poder informativo del Estado que abarque todos los órganos de contralor público, sociedades civiles y emisoras públicas y privadas como se despliega con más detalle en la siguiente sección.

# 7. Revolución digital y la constitución de un poder informativo del Estado

La revolución digital tuvo y tiene consecuencias en toda la sociedad puesto que un Ciudadano común con escasos conocimientos de computación es capaz de generar información y convertirse en una usina informacional, este empoderamiento digital del Ciudadano produjo un incremento exponencial de la cantidad de información generada en la sociedad, multiplicidad de canales y fuentes de información en las cuales las estructuras pre-digitales del Estado no se encontraban preparadas para asimilar.

Así como los partidos políticos son los encargados de trasvasar esa producción informacional de la ciudadanía en la forma de proyectos e ideas en el propio Estado, una función importante del mismo era históricamente atribuida al periodismo, donde se canalizaban parte importante de la producción informacional de sociedad y contribuía al debate público, lo que produjo ganancias de tiempo para toda la sociedad al acelerar los proceso de construcción de consensos en la espiralización del silencio. Si bien el periodismo es sido considerado como un cuarto poder no estructurado, las organizaciones civiles de contralor, la ciudadanía en internet también son considerados poderes no estructurados y también contribuyen a la construcción de consensos Ciudadanos.

El empoderamiento digital de la ciudadanía a través de internet es quizás las estructuras de servopoder Ciudadano que más impulsó la revolución digital, pues a través de los medios tecnológicos que el Ciudadano común, ahora es capaz de acaparar funciones que por razones técnicas solo eran atribuidas al periodismo. Las organizaciones civiles también son beneficiadas con las nuevas tecnologías lo que le permite al Ciudadano apartidario desarrollar funciones de producción periodística para el contralor civil, participación en la gobernanza cuando los gobiernos son abiertos, etc., funciones cívicas hace un siglo en ese tiempo monopolizadas exclusivamente por la prensa.

Este desplazamiento funcional del empoderamiento digital de la ciudadanía informatizada no solo genera un impacto en los poderes (des)estructurados del Estado sino que también propugna una crisis en los otros poderes no estructurados de la sociedad lo que lleva al incremento de la presión informacional sobre la sociedad toda, generados estos por los esfuerzos de cada parte por retener la mayor cantidad de poder posible a costa de perdida de tiempo social de la ciudadanía perturbada y desinformada.

Ahora con la irrupción de la revolución digital hubo una disfuncionalidad de los poderes no estructurados, así como aumento el de la ciudadanía también lo hizo el de los medios concentrados los cuales se ven motivados a concentrar aún más poder de emisión y presión informática para no perder su rol y capacidad de influencia informativa en su papel de encodificador y regulador de masas en la autoregulación cibernética informacional de la sociedad. Estas disfuncionalidades abren la puerta a la necesidad de concertación de nuevas democracias deliberativas digitales de mayor interacción informacional entre el Estado y sus Ciudadanos puesto que las tecnologías así ahora lo permiten.

Por otra parte la lucha friccional entre las emisoras tradicionales para mantener su poder desestructurado genera entropía e inflación informativa espúrea puesto que para mantener su preponderancia incrementan la presión informacional con tal de no percibir una disminución en el impacto de sus emisiones y ediciones. Esta inflación reduce el valor útil de la información volviéndola espúrea haciendo que el demandante de información pierda tiempo de contemplación y recursos en la observación de ruido perturbante en lugar de información verosímil.

Este avance de la presión informativa desde un sector de los medios masivos puede verse como una transición incompleta de los Estados hacia verdaderos Estados digitales se ve impedida de serlo debido no solo a la revolución informática sino a que el *funcionamiento* y las *funciones* del Estado que requieren un mayor aislamiento y procesos para accionar se vean imposibilitados de hacerlo al encontrarse informativamente aturdidos desembocando en pérdidas de tiempo irreversible para la sociedad toda.

En términos de la ecuación de Klitgaard expandida relacional –*i. e.* **ecuación 5**–, se aprecia el fenómeno no necesariamente sobre la corrupción, pero si también directamente sobre el capital social como se observa en la **ecuación** 6, donde la influencia informativa-comunicacional sobre el capital societario resulta fundamental para sostener su densidad y cohesión. Las fricciones entre poderes no estructurados también afectan ese componente generando un incremento en el ruido informacional lo que repercute negativamente en su densidad y en pérdidas irrecuperables de energía humana y tiempo para la sociedad al estar absorbida por recursos informacionales ficticios. Por estos motivos, surge la necesidad de nuevos mecanismos de regulación informacional que además de defender la libertad de expresión pueda ser capaz de acolchonar la presión informacional dado que cuando la presión informativa (PI) supera cierto umbral después del cual, al romperse los servomecanismos de equilibrio de retroalimentación comunicacional, los flujos informativos societales se vuelven caóticos afectando directamente a la densidad del capital social y por ende el bienestar social agregado.

Una forma de reducción de ese ruido puede ser el de la estructuración un poder informativo de Estado legitimador de la ciudadanía digital como garante de la neutralidad de la red, la libertad de expresión digital, la universalidad de la conectividad, multipolaridad de voces; por medios humanos o artificiales podría monitorearse la entropía de la suma agregada de las emisoras, sin ánimo de censura alguna pero si con la capacidad de calificar la/s misma/s como fuente de perturbación social pues los mismos anulan la capacidad planificadora y afectan la sensación y el tejido social contribuyendo al desglose del capital societario.

Esta institucionalización en un poder orgánico informativo del Estado hace a la necesidad de observación y representación de los medios de comunicación tradicionales, incluyendo las sociedades civiles y toda usina emisora con entidad desde la ciudadanía digital para asegurar la universalización de la conectividad, la neutralidad de la red, la pluralidad de voces, un periodismo digital ético, la participación ciudadana y la renovación del poder dentro del mismo así como de la necesidad de un periodismo digital crítico pero también ético que no contribuya a la naturalización del comportamiento corrupto en la sociedad sino que lo combata efectivamente promoviendo la ejemplaridad de Ciudadanos probos.

Por otra parte, así como toda labor investigativa de la sociedad que necesite de cierto aislamiento, entre ellos los fallos y procesos judiciales, son a la vez fuentes y objeto de presión informativa, superado un umbral, el ruido producto de la primera parece tener impacto sobre la velocidad de los procesos de determinadas causas respecto de otras, un poder informacional activo podría intervenir asegurando la homogeneidad y estandarización de los procesos cuando esto contribuyen a incrementar la presión informativa. Un hipotético poder informativo podría, además de considerar a la información como un recurso social necesario también considere al

tiempo –la pérdida del mismo– como un recurso humano preciado no renovable con que cuenta toda sociedad, así como con capacidad de intervención Estatal sobre las funciones del resto de los poderes estructurados y desestructurados, –e. g. en casos de desnutrición o tenga poder de gestión sobre el ejecutivo, operaciones informacionales sobre causas judiciales, etc.–

Un poder informativo del Estado, que sea activado de forma no permanente sobrepasarse cierto umbral entrópico que determine que el ruido comunicacional afecta las funciones de los poderes estructurados del Estado al afectar la encodificación legislativa y la decodificación judicial de las normas, códigos y leyes de la república. Esta interferencia informativa solo puede darse cuando existe una coordinación desinformativa metodológicamente reactiva, una interferencia sistemática desde donde se provoca dicha entropía, que afecta directamente los equilibrios informativos de la sociedad.

La activación de un poder informativo republicano debería poseer una metodología para determinar su momento en el tiempo para su actuación y funcionamiento, a partir de niveles de entropía de la información emitidos por fuentes comunicacionales masivas donde el ruido y la interferencia sistemática es capaz de hendir los servomecanismos de auto-regulación societaria comunicacional, cuando la dispersión de los mensajes encodificados con los decodificados por las masas. conformado por oficiales con poder público pero que también necesitan instrumentos informáticos de lógica artificial capaces de ser configurados para detectar umbrales de ruido societario que auxilien al factor humano en la regulación comunicacional y la aseguranza del bienestar.

## 8. Algunas conclusiones

Se presentó una actualización de la ecuación de Klitgaard donde se intentó agiornar esta útil y pedagógica –a la vez que simple– explicación de corrupción por parte de aquel autor, ahora incorporando conceptos provenientes teoría de la información, teoría de la comunicación y la importancia del capital social en cualquier sociedad toda. Asumiendo la existencia de equilibrios éticos hipotéticos de acuerdo a dos contextos informacionales particulares correspondientes a la presencia de mecanismos informativos disímiles, se experimento ejemplificativamente con el accionar corruptor o ejemplificativo de dos funcionarios públicos pertenecientes a una misma sociedad.

Por otra parte, además de brindar mejoras en el entendimiento y la representación del fenómeno corrupto, se desplegó una breve crítica a los índices de percepción de la corrupción, lo que pudiera servir como punto de partida para empezar a discutir nuevos enfoques para la resolución de su problemática. La mejor forma de lucha contra el flagelo, como sugiriera Tanzi [22], es la voluntad política, o según Cortina [5] una ciudadanía moral comprometida; sin embargo, en contextos de desarrollo donde una sociedad atraviesa necesidades básicas, fisiológicas insatisfechas la discusión pública del problema de la corrupción puede verse limitado o restringido por otras agendas informáticas pudiendo estar bien o no apoyado en cuanto las usinas desinformacionales o informacionales auténticas de la sociedad territorialmente situada.

Esto determinará si el equilibrio ético prevaleciente en ella sea corruptivo o ejemplificativo. Tanto la ecuación de Klitgaard actualizada como la existencia de un equilibrio corruptor o ejemplificativo hipotético, aunque resultan útiles como herramientas conceptuales que ayudan a comprender el fenómeno, distan y lo hacen mucho de la promulgación y desarrollo de una filosofía moral comprometida en la construcción de una sociedad civil ética que se sienta insustituible en su acción en su tiempo y sus iguales para forjar el camino a lo que Cortina entendía como el trajín desde una *ciudadanía política* hacia una *ciudadanía moral* (Cortina, [5]).

Una opción es la de considerar una solución sistémica a esta problemática de índole compleja incluyendo el accionar ético del Ciudadano dentro de su contexto. El abordaje sistémico de la corrupción no quita responsabilidad a las personas, este las incluye en el análisis de las condiciones externas que influyen en las personas que realizan prácticas corruptas al considerar el tiempo de vida de cada individuo como parte de una vida colectiva (Gomez y Bello, [10]), por lo que la actuación ejemplar o corrupta de los mismos repercute positivamente dentro del capital social al que estos mismos pertenecen.

Cuanto mayor sea la responsabilidad institucional mayor debe ser la transparencia, resultando una relación inversamente proporcional en lo que hace a la relación entre responsabilidad republicana y privacidad, puesto no solo por el necesario contralor civil de los Ciudadanos probos sino también como una demanda necesaria para que los canales informativos entre el Estado y sus Ciudadanos que lo componen no sea perturbada ni distorsionada en favor de ninguna clase. Afortunadamente el discernimiento y las capacidades informativas de la población cívicamente formada es cada vez mayor y los efectos desmedidos de emisoras desinformantes éticamente irresponsables se reflejan menos en víctimas humanas y más en angustia generalizada, malestar o pérdida de público receptor cautivo.

Empezar a discutirse la necesidad de un modelo alternativo al que se propone con el intento de atacar al flagelo corruptivo más allá de la observación de índices de percepción unidimensional para enfocar los esfuerzos en una problemática informacional del fenómeno y la información como servomecanismo de auto-regulación societal. Los perjuicios de la corrupción estructural profunda se observa también en la ruptura del capital societario mediante ineficacia en la provisión de servicios públicos, el desconocimiento de los derechos cívicos, etc. Medidas de política pública concretas para el fortalecimiento del capital social, universalización de la conectividad, neutralidad de la red, digitalización de los organismos anti-corrupción que aseguren pormenorizadamente en todo momento los estados y el seguimiento de los patrimonios de los representantes oficiales principales de cada nación, etc. no pueden dejar mencionarse así como constituyen las mejores opciones para una lucha que se da ejemplificativamente y necesita tiempo.

La reducción de las brechas digitales, la neutralidad efectiva de la red podrían ser mejor relevados a través del diseño de índices, que en lugar de basarse en conceptos subjetivos, reflejen la labor ejemplificadora de la ciudadanía así como las posibilidades de acceso efectivo a la información pública, la capacidad de transformación socialcultural del Ciudadano a través de medios digitales, de participación efectiva de la ciudadanía en el contralor público por parte de los Ciudadanos, universalidad de la conectividad de la población, etc. Así como lo es la necesaria formación cívica crítica de la ciudadanía, es igual de necesaria la formación y el desarrollo digital de la ciudadanía en todos los aspectos.

Por otra parte las asociaciones que luchan contra el flagelo de la corrupción que a su vez tienen la responsabilidad de ser formadoras de opinión deberían tener políticas interna de doble transparencia y probidad en cuanto a sus fuentes financiamiento, –*i. e.* presentar en donde se determine una lista institucional detallada tanto nominal como monetaria de los contribuyentes–, donde a mayor atomización y diversidad de los aportes percibidos mayor será la probidad de toda asociación con capacidad de formar opinión sobre otros, su accionar debe ser ejemplar, ofreciendo transparencia completa en toda su estructura informacional y la mayor objetividad posible puesto que estas mismas asociaciones son también las demandantes de data institucional de todo tipo, desde organismos, sociedades y hasta de personas físicas. La organicidad y estructuración de un poder informativo podría ayudar a acometer esta tarea con nuevas metodologías dirigidas hacia problemas complejos.

Es lógico pensar que en sociedades más avanzadas que otras en cuanto que priman una mayor demanda de necesidades de satisfactores más elevadas que básicas, y cuenten con el tiempo la energía y la motivación para acometer con la tarea tenderán a ser –a los ojos de sociedades más adelantadas– más estructuralmente corruptas que otras; de cualquier forma, las usinas mediáticas masivas seguirán teniendo una función moral tanto en la formación de la opinión pública como en amplificar o reducir el accionar comunicativo de la ciudadanía mientras no surja una verdadera proto-democracia deliberativa digital que las reemplace o al menos la complemente desde una participación directa o cuasi directa a través de nuevos medios tecnológicos.

Además de la impronta ética augurada por Cortina, los nuevos contextos democráticos digitales debe asegurarse el paso hacia una verdadera ciudadanía moral en contextos informacionales cada vez más complejos, para lo cual además de la irreemplazable acción ciudadana hace falta una nueva generación de derechos humanos de índole informática y herramientas computacionales impostergables que así lo aseguren. Esto representa otra oportunidad para utilizar los conocimientos de las tecnologías de la información en favor del desarrollo de nuevas herramientas para la ciudadanía, que aseguren el desarrollo de instrumentos digitales mínimos sobre aquellas capas de información cívica necesarias para el contralor civil de los Ciudadanos digitalmente formados.

El desarrollo de software y/o procesos de explotación de la información que puedan coadyuvar a un mayor control de los procesos cívicos por parte de la ciudadanía y las instituciones de contralor públicas y privadas se vuelve una oportunidad digna de esfuerzo, considerando la formación digital natural de las generaciones presentes y por venir. El futuro desarrollo de herramientas avanzadas de contralor supondrán un paso más hacia mejores y más maduros métodos de gobernanza abierta y participación ciudadana para la conformación de una nueva democracia informativa deliberativa.

En escenarios de corrupción estructural o de fuerte concentración mediática cuando la participación ciudadana se vea anonadada y desinteresada éticamente en tomar parte, así como la presión informativa cruce cierto umbral de caoscidad donde fallen los servomecanismos informacionales, esta puede elucidarse a través de poderes informativos capaces de calificar las fuentes informativas como legítimas y disruptivas que hagan a la necesidad de intervención ciudadana, un camino posible de tal legitimación comprende la conformación orgánica de un cuarto poder republicano que acapare las dimensiones informativas representativa de poderes hasta ahora no estructurados.

# Referencias

**RESUMEN**

Se configura una actualización de la ecuación de Klitgaard para la explicación completa del fenómeno corrupto en escenarios informativos y comunicacionales complejos. Incorporando conceptos de teoría del capital social, se desliza la existencia de equilibrios éticos corruptivos o ejemplificativos acordes a la suposición de distintos servomecanismos y entornos comunicacionales aparentes. Posteriormente se discute la (in)eficacia de los índices de percepción como punto de partida para la elaboración de alternativas sistémicas reales y globales en la lucha anti-corrupción así como de instauración de un poder informativo del Estado para la reducción de la entropía y la presión informativa societal.

**Palabras Clave:** sistemas complejos; teoría de la corrupción; teoría de la información; mecanismos informativos; equilibrio ético corruptivo-ejemplificativo; índices de percepción; corrupción informativa; poder informativo

**RESUMO**

Configura-se uma atualização da equação de Klitgaard para a explicação completa do fenômeno corrupto em cenários informativos e comunicativos complexos. Incorporando conceitos de teoria do capital social, desliza a existência de equilíbrios éticos corruptivos ou ejemplificativos de acordo com a suposição de diferentes servomecanismos e ambientes de comunicação aparentes. Posteriormente, discute-se a (in)eficácia dos índices de percepção como ponto de partida para a elaboração de alternativas sistêmicas reais e globais na luta anti-corrupção, bem como de instauração de um poder informativo do Estado para a redução da entropia e a pressão informativa societal.

**Palavras-Chave:** sistemas complexos; teoria da corrupção; teoria da informação; mecanismos informativos; equilíbrio ético corruptivo-ejemplificativo; índices de percepção; corrupção informativa; poder informativo

# Distribución semántica

Fig. 2: Distribución semántica

Este trabajo adhiere a la campaña global de Naciones Unidas contra la corrupción.
http://www.anticorruptionday.org/